\documentclass[%
 aip,
 amsmath,amssymb,
 reprint,%
]{revtex4-1}

\usepackage{graphicx}
\usepackage{dcolumn}
\usepackage{bm}
\usepackage[per-mode=symbol]{siunitx}
\usepackage[utf8]{inputenc}
\usepackage[T1]{fontenc}
\usepackage{mathptmx}
\usepackage{etoolbox}
\usepackage{ragged2e}
\usepackage[colorlinks=true,breaklinks=true,allcolors=blue]{hyperref}
\usepackage[capitalise,nameinlink]{cleveref}
\usepackage{soul}
\usepackage[all]{nowidow}
\makeatletter
\def\@email#1#2{%
 \endgroup
 \patchcmd{\titleblock@produce}
  {\frontmatter@RRAPformat}
  {\frontmatter@RRAPformat{\produce@RRAP{*#1\href{mailto:#2}{#2}}}\frontmatter@RRAPformat}
  {}{}
}%
\makeatother
\begin{document}

\preprint{AIP/123-QED}

\title[Analog programming of Al$_2$O$_3$/TiO$_\textrm{2-x}$ memristor at \SI{4.2}{\kelvin} after metal-insulator transition suppression by cryogenic reforming]{Analog programming of CMOS-compatible Al$_2$O$_3$/TiO$_\textrm{2-x}$ memristor at \SI{4.2}{\kelvin} after metal-insulator transition suppression by cryogenic reforming}
\author{Pierre-Antoine~Mouny$^*$}
\affiliation{Institut Interdisciplinaire d’Innovation Technologique (3IT), Université de Sherbrooke, Sherbrooke, Québec J1K 0A5, Canada}
\affiliation{Laboratoire Nanotechnologies Nanosystèmes (LN2) – CNRS IRL-3463 – 3IT, Sherbrooke, Québec J1K 0A5, Canada}
\affiliation{Institut Quantique (IQ), Université de Sherbrooke, Sherbrooke, Québec J1K 2R1, Canada}
\email[Corresponding author: ]{pierre-antoine.mouny@usherbrooke.ca}
\author{Raphaël~Dawant}
\author{Bastien~Galaup}
\affiliation{Institut Interdisciplinaire d’Innovation Technologique (3IT), Université de Sherbrooke, Sherbrooke, Québec J1K 0A5, Canada}
\affiliation{Laboratoire Nanotechnologies Nanosystèmes (LN2) – CNRS IRL-3463 – 3IT, Sherbrooke, Québec J1K 0A5, Canada}
\author{Serge~Ecoffey}
\affiliation{Institut Interdisciplinaire d’Innovation Technologique (3IT), Université de Sherbrooke, Sherbrooke, Québec J1K 0A5, Canada}
\affiliation{Laboratoire Nanotechnologies Nanosystèmes (LN2) – CNRS IRL-3463 – 3IT, Sherbrooke, Québec J1K 0A5, Canada}
\author{Michel~Pioro-Ladrière}
\affiliation{Laboratoire Nanotechnologies Nanosystèmes (LN2) – CNRS IRL-3463 – 3IT, Sherbrooke, Québec J1K 0A5, Canada}
\affiliation{Institut Quantique (IQ), Université de Sherbrooke, Sherbrooke, Québec J1K 2R1, Canada}
\author{Yann~Beilliard}
\author{Dominique~Drouin}
\affiliation{Institut Interdisciplinaire d’Innovation Technologique (3IT), Université de Sherbrooke, Sherbrooke, Québec J1K 0A5, Canada}
\affiliation{Laboratoire Nanotechnologies Nanosystèmes (LN2) – CNRS IRL-3463 – 3IT, Sherbrooke, Québec J1K 0A5, Canada}
\affiliation{Institut Quantique (IQ), Université de Sherbrooke, Sherbrooke, Québec J1K 2R1, Canada}

\date{\today}

\begin{abstract}
The exploration of memristors' behavior at cryogenic temperatures has become crucial due to the growing interest in quantum computing and cryogenic electronics. In this context, our study focuses on the characterization at cryogenic temperatures (\SI{4.2}{\kelvin}) of TiO$_\textrm{2-x}$-based memristors fabricated with a CMOS-compatible etch-back process. We demonstrate a so-called cryogenic reforming (CR) technique performed at \SI{4.2}{\kelvin} to overcome the well-known metal-insulator transition (MIT) which limits the analog behavior of memristors at low temperatures. This cryogenic reforming process was found to be reproducible and led to a durable suppression of the MIT. This process allowed to reduce by approximately 20\% the voltages required to perform DC resistive switching at \SI{4.2}{\kelvin}. Additionally, conduction mechanism studies of memristors before and after cryogenic reforming from \SI{4.2}{\kelvin} to \SI{300}{\kelvin} revealed different behaviors above \SI{100}{\kelvin}, indicating a potential change in the conductive filament stoichiometry. The reformed devices exhibit a  conductance level that is 50 times higher than ambient-formed memristor, and the conduction drop between \SI{300}{\kelvin} and \SI{4.2}{\kelvin} is 100 times smaller, indicating the effectiveness of the reforming process. More importantly, CR enables analog programming at \SI{4.2}{\kelvin} with typical read voltages. Suppressing the MIT improved the analog switching dynamics of the memristor leading to approximately 250\% larger on/off ratios during long-term depression (LTD)/long-term potentiation (LTP) resistance tuning. This enhancement opens up the possibility of using TiO$_{\textrm{2-x}}$-based memristors to be used as synapses in neuromorphic computing at cryogenic temperatures.
\end{abstract}

\maketitle

Memristors are among the leading candidates for the large-scale integration of non-volatile memory and artificial synapses in neuromorphic architectures \cite{Li2018}. While their characteristics have been widely investigated at \SI{300}{K}, the growing interest in quantum computing and cryogenic electronics has led to new concepts based on memristors to overcome some of the engineering challenges encountered in scaling-up quantum computers \cite{Vandersypen2017}. Recent works propose developing cryogenic-compatible CMOS-memristor hybrid circuits for applications such as tunable cryogenic DC sources \cite{Mouny_2023}, the charge state autotuning of quantum dots \cite{Czischek2021}, quantum error correction \cite{Marcotte2023} or spiking neural networks \cite{Chen_spiking}. In this context, cryogenic studies have been conducted on various types of emerging memory based on HfO$_2$\cite{Voronkovskii2019, Fang2015, Cryo_HfO2, Blonkowski2015}, AlO$_\textrm{x}$ \cite{AlOx}, NiO$_\textrm{x}$ \cite{Alagoz2023}, TaO$_\textrm{x}$ \cite{Zhang2014}, TiO$_\textrm{2-x}$ \cite{, Pickett2011, Beilliard2020} and ferroelectric HZO \cite{manchon2022study, HZO_cryo}. Among these candidates, TiO$_{\textrm{2-x}}$ based memristors present the most promising characteristics at room temperature including fully analog behaviour, low switching voltages and CMOS-compatibility \cite{Kim2021}. However, the cryogenic DC characterizations of memristors based on TiO$_{\textrm{2-x}}$ exhibit  a metal-insulator transition (MIT) at low voltages below \SI{125}{\kelvin}\cite{Beilliard2020, Pickett2011, Alagoz2019}. This MIT is attributed to the Ti$_4$O$_7$ Magnéli phase and hinders its usage at cryogenic temperatures by limiting its ohmic region.\linebreak

In this work, we report a specific reforming methodology at cryogenic temperatures aiming to suppress the MIT of Al$_2$O$_3$/TiO$_\textrm{2-x}$ memristors. We investigate the impact of both ambient temperature forming (AF) and cryogenic reforming (CR) on the conduction mechanisms of our devices from \SI{4.2}{\kelvin} to \SI{300}{\kelvin}. Finally, we demonstrate that the suppression of the MIT allows the quasi-linear analog tuning of the memristor resistance at \SI{4.2}{\kelvin} using voltage pulses in a long-term depression (LTD)/long-term potentiation (LTP) tuning scheme. \linebreak

\begin{figure*}

  \includegraphics[width=0.95\linewidth]{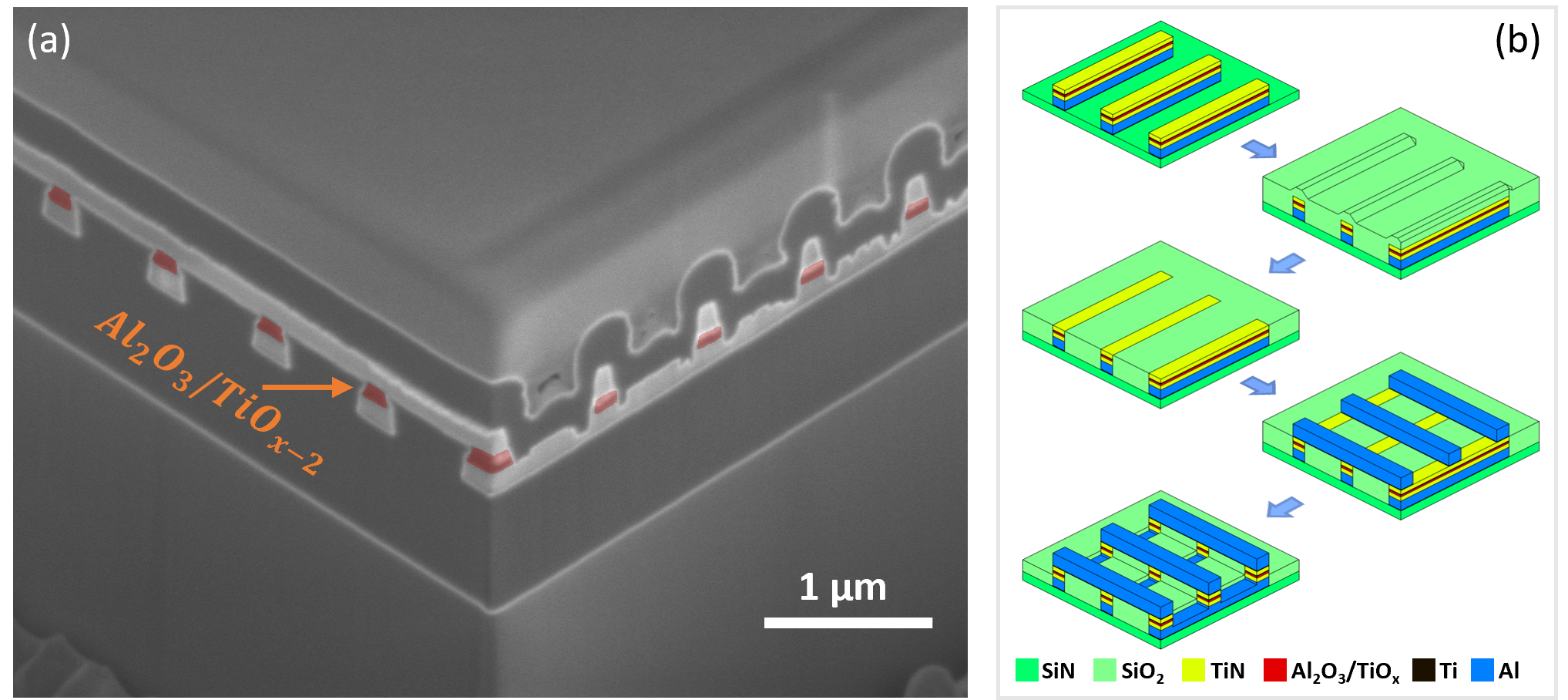} 
\vfill 
  \includegraphics[width=1\linewidth]{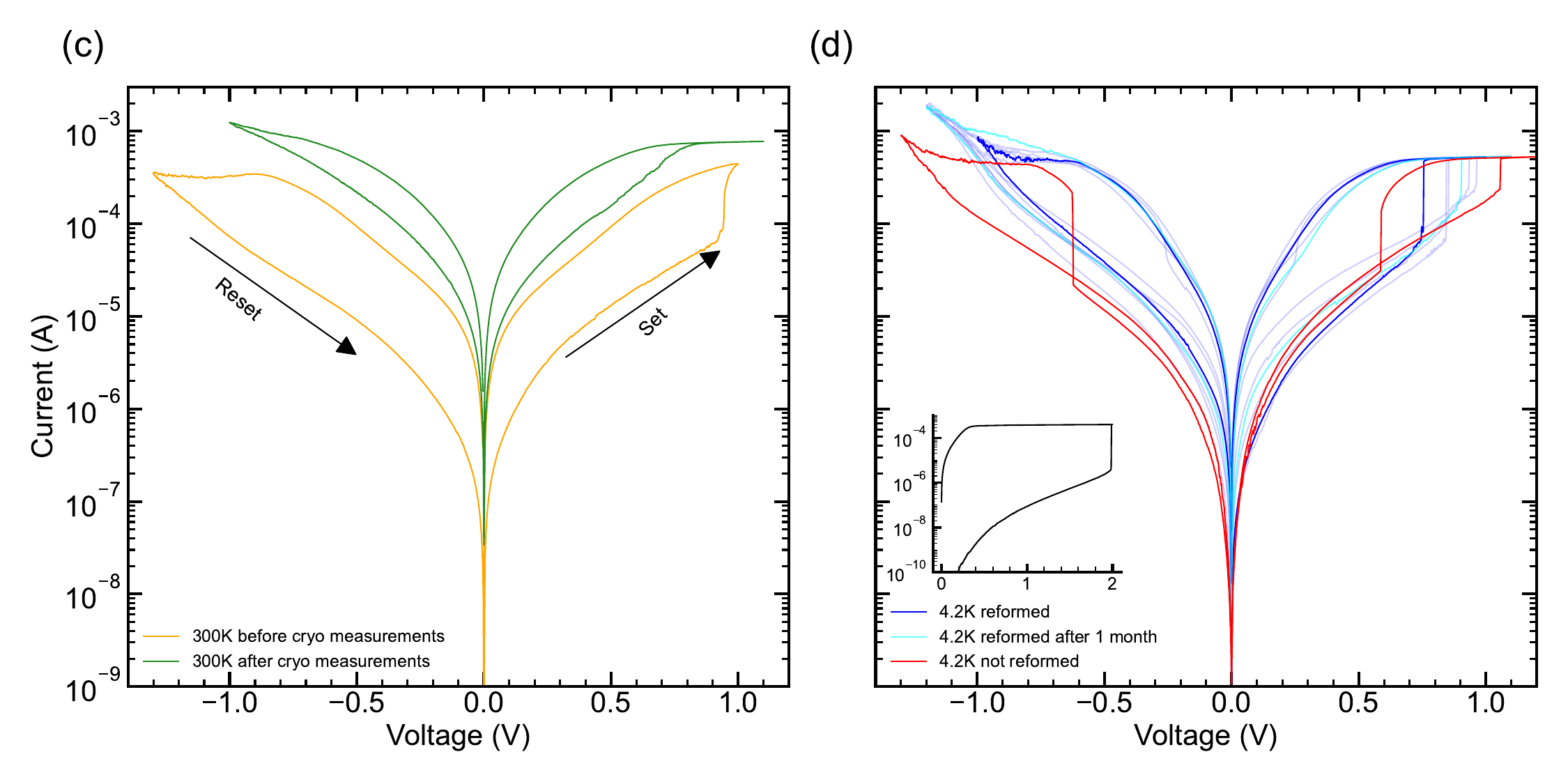}
\caption{\label{fig:DC} (a) Focused ion beam (FIB) isometric cross-section of the devices. The Al$_2$O$_3$/TiO$_{\textrm{2-x}}$ crosspoints patterned at the intersection of the bottom electrodes (BEs) and top electrodes (TEs) are highlighted in red. (b) Schematic representation of the fabrication process flow: The BE and the active stack are patterned and etched together, followed by the deposition of an SiO$_2$ layer. The structures are revealed by chemical-mechanical polishing (CMP) and the TEs are patterned similarly to the BEs. During this step, the remaining active stack between the TEs is etched with the TEs. (c) I-V characteristics of the Al$_2$O$_3$/TiO$_{\textrm{2-x}}$ crosspoints at \SI{300}{\kelvin} before and after cryo-reforming. (d)
I-V characteristics of ambient-formed (AF) and cryo-reformed (CR) memristors at \SI{4.2}{\kelvin}. The reforming process is shown in the inset. Each shadowed blue curve represents a different CR memristor, highlighting the device-to-device variability. A CR memristor was measured at \SI{4.2}{\kelvin} one month after cryo-reforming to show the retention of the MIT suppression on the cyan curve.}
\end{figure*}

The devices investigated consist of Ti/Al/TiN - Al$_2$O$3$/TiO$_\textrm{2-x}$/Ti - TiN/Ti/Al, which represent the bottom electrode (BE), active stack and top electrode (TE), respectively. The process flow is illustrated in \cref{fig:DC}(b). The first step of fabrication involves the sputtering deposition of a Ti (\SI{10}{\nano\metre})/Al (\SI{150}{\nano\metre})/TiN (\SI{70}{\nano\metre}) stack for the BE. This is followed by the deposition of the active switching stack of Al$_2$O$_3$ (\SI{1.4}{\nano\metre}) through atomic layer deposition, and TiO$_\textrm{2-x}$ (\SI{30}{\nano\metre}) by reactive sputtering, along with Ti (\SI{10}{\nano\metre})/TiN (\SI{70}{\nano\metre}). The vacuum was not broken between the depositions of the TiO$_\textrm{2-x}$, Ti and TiN layers. A \SI{200}{\nano\metre}-wide BE is patterned using electron beam lithography (EBPG5200), a negative resist (MaN2405) and an inductively coupled plasma (ICP) etching process with BCl$_3$/Cl$_2$/Ar chemistry. The BEs and the switching stack are both encapsulated through the deposition of \SI{500}{\nano\metre} of SiO$_2$ using plasma-enhanced chemical vapor deposition (PECVD). Subsequently, a chemical-mechanical polishing (CMP) process is applied to planarize the SiO$_2$ surface and reveal the top surface of the BE. The TiN layer serves as a polishing stop layer for the CMP process. The SiO$_2$ dishing between the BE, measured by atomic force microscopy (AFM), is less than \SI{5}{\nano\metre}. A \SI{200}{\nano\metre}-wide TE made of Ti (\SI{10}{\nano\metre})/Al (\SI{400}{\nano\metre}) is deposited and patterned similarly to the BE. The active stack between the TE lines is removed during the etching step of the TE. The devices are then encapsulated with a \SI{600}{\nano\metre} layer of SiO$_2$ deposited by PECVD. Vias to contact the electrodes are patterned using electron beam lithography and ICP etching with SF$_6$ chemistry. Finally, Al (\SI{400}{\nano\metre}) connection lines are patterned to address the memristors. After this, a 1-minute rapid thermal annealing step at \SI{350}{\celsius} under an N$_2$ atmosphere is performed. As shown in the focused ion beam (FIB) isometric cross-section in \cref{fig:DC}(a), the memory stack lies only at the intersection of the BEs and TEs, like a similar approach presented in the literature \cite{Dawant2022}.\linebreak

The memristor crosspoints are characterized from \SI{4.2}{\kelvin} to \SI{300}{\kelvin} with a Keysight B1500A semiconductor device parameter analyzer with a  \SI{200}{\mega samples/\second} waveform generator module (WGFMU) on a Lakeshore CPX-VF cryogenic probe station.
Both the sample and the magnet  stage are heated when the device is warmed up to ensure temperature stability. For all measurements, the BEs are grounded and the signals are applied to the TEs. The memristor resistances are measured at low bias (\SI{0.02}{\volt}) with a DC sweep for the conduction mechanism studies, and they are measured with a \SI{0.2}{\volt}/\SI{10}{\micro\second} read pulse during pulsed characterizations. For all DC measurements, we limited the maximum current through the memristor to \SI{500}{\micro\ampere} using a custom current compliance board placed at \SI{300}{\kelvin} to avoid overshoot during forming and DC sweeps. Prior to the cooldown, 10 memristors are ambient-formed (AF) at \SI{300}{\kelvin}. We perform I-V measurements to benchmark the non-volatile switching of our devices at room temperature (see \cref{fig:DC}(c)). The sample is then cooled down to \SI{4.2}{\kelvin}.\linebreak

The first I-V measurements at \SI{4.2}{\kelvin} on an AF memristor exhibit the MIT introduced by the Ti$_4$O$_7$ Magnéli phase, as depicted by the red curve in \cref{fig:DC}(d) \cite{Pickett2011, Beilliard2020}. By applying an approximately two-times-larger positive DC sweep $V_{\textrm{SET}}=$ \SI{2}{\volt} with a \SI{500}{\micro\ampere} compliance current to an AF memristor, one can observe an abrupt increase of the measured current up to the compliance current at $\approx$\SI{2}{\volt}, as shown in the inset of \cref{fig:DC}(d). This phenomenon is similar to electroforming and suggests that the conductive filament of the memristor is reformed. After this CR, we observe that consecutive DC switchings do not exhibit the characteristic MIT anymore as shown by the blue curve in \cref{fig:DC}(d). The CR process is reproducible as we create five reformed devices with similar DC behavior. Moreover, the DC cycle at \SI{300}{\kelvin} after the cooldown exhibits a lower low resistance state (LRS) and high resistance state (HRS) compared to before CR was performed, suggesting that the diameter of the conductive filament is larger due to CR as observed in \cref{fig:DC}(c). These CR memristor DC cycles show butterfly-shaped switchings as the room-temperature I-V curves measured prior to the cooldown. However, one can note that lower $V_{\textrm{SET}}$ and $V_{\textrm{RESET}}$ were needed to switch the memristor at \SI{4.2}{\kelvin} which is an interesting feature for low-power applications. We also investigate the retention of the MIT suppression by cooling down and measuring the same CR device one month later. The I-V characteristics of the CR memristor remained without MIT at \SI{4.2}{\kelvin} (see the cyan curve in \cref{fig:DC}(d)). It suggests that a permanent change to the conductive filament was made through CR yielding to different conduction mechanisms at cryogenic temperatures.\linebreak

\begin{figure}
\begin{minipage}[b]{1\columnwidth}
  \includegraphics[width=1.05\linewidth]{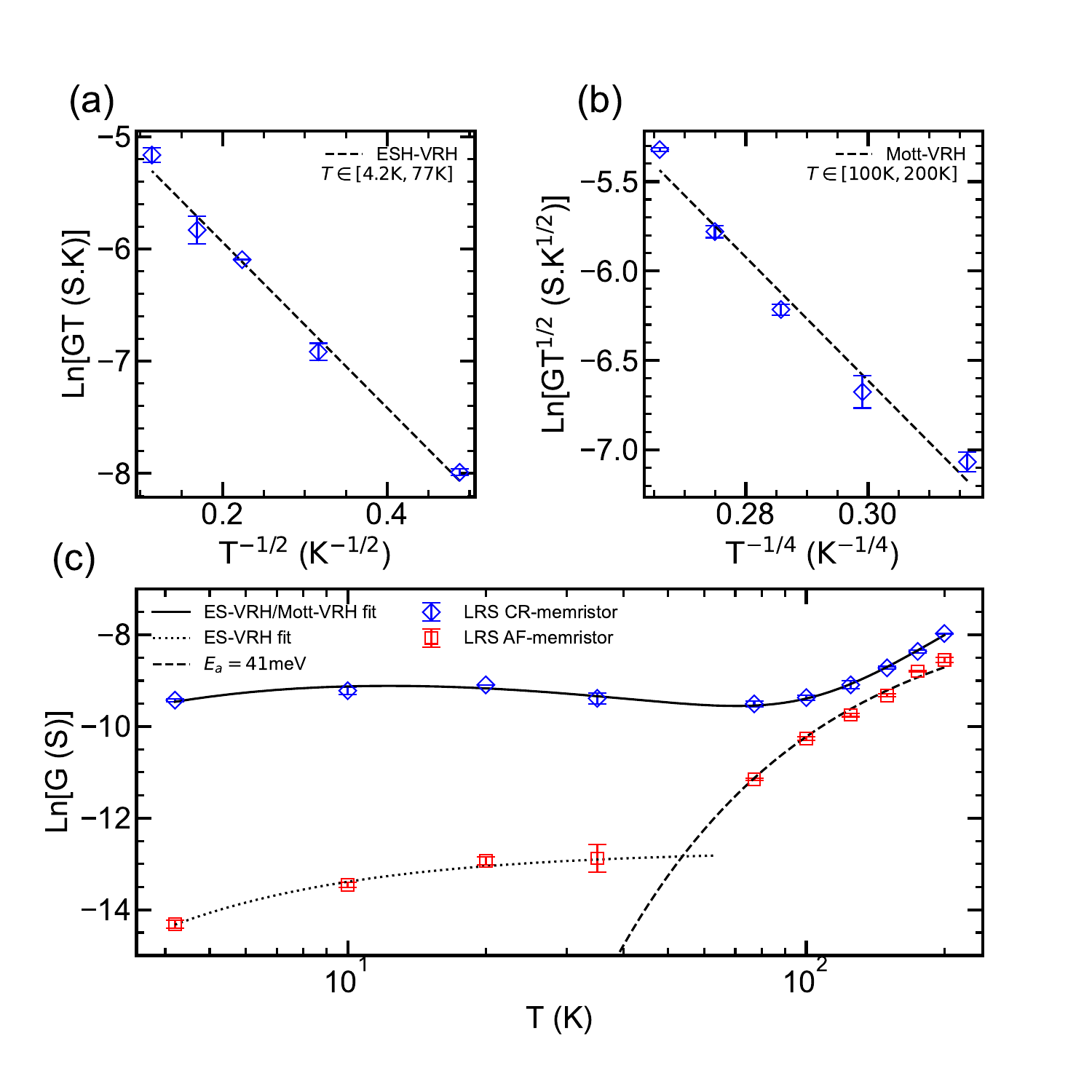}
  \caption{\label{fig:conduct} (a)  Efros–Shklovskii VRH conductance fit (\cref{eq:ES_eq}) for a CR memristor at LRS between \SI{4.2}{\kelvin} and \SI{77}{\kelvin}. (b)  Mott VRH conductance fit (\cref{eq:Mott_eq}) for a CR memristor at the LRS between \SI{100}{\kelvin} and \SI{200}{\kelvin}. (c) Temperature dependence of the memristor conductance. The CR memristor LRS conductance is fitted by \cref{eq:global_conduct} using the parameters from \cref{tab:fit_params} while the AF memristor LRS conductance is fitted with an Arrhenius law (\cref{eq:conduct} $\alpha=1$) for $T>$\SI{35}{\kelvin} and an ES-VRH law for $T\leq$\SI{35}{\kelvin}. The conductances are measured at \SI{20}{\milli\volt}. The AF memristor and CR memristor conductances are measured five times after a RESET/SET operation at each temperature to assess the cycle-to-cycle variability shown in the error bars.}
\end{minipage}\hfill 
\end{figure}

To investigate the change in the conduction mechanisms involved in the MIT suppression,  we perform additional DC cycle measurements for both AF and CR devices from \SI{10}{\kelvin} to \SI{200}{\kelvin} and extract the conductances at each temperature for the LRS by reading the current at \SI{20}{\milli\volt}. As shown in \cref{fig:conduct}(c), the conductance decreases with temperature for the two types of devices. For the TiO$_{\textrm{2-x}}$ memristors exhibiting the MIT (AF memristors), the temperature-dependant conductance $G(T)$ can be described by an Arrhenius equation:
\begin{equation}\label{eq:conduct}
    G^{\textrm{MIT}}(T)=G_0\exp\left(-\frac{E_a}{k_BT}\right)^{\alpha}
\end{equation}
where $\alpha$ depends on the transport mechanism (e.g., $\alpha=1$ for nearest neighbor hopping (NNH) conduction \cite{mott2012electronic}), $G_0$ is a pre-exponential factor and $E_a$ is the thermal activation energy \cite{mott2012electronic, Beilliard2020}. The LRS conductance of the AF memristor fits linearly with \eqref{eq:conduct} for $\alpha = 1$ between \SI{300}{\kelvin} and \SI{77}{\kelvin} and $\alpha=0.5$ between \SI{77}{\kelvin} and \SI{4.2}{\kelvin} which corresponds to Efros–Shklovskii variable-range
hopping (ES-VRH) as reported in Ref.\cite{Beilliard2020}.\linebreak

Regarding the CR devices, \cref{fig:conduct}(b) shows a linear relationship between $\ln \left(GT^{1/2}\right)$ and $T^{-1/4}$ from \SI{100}{\kelvin} to \SI{200}{\kelvin}. This indicates that the conduction in the oxide layer is governed by Mott's variable-range-hopping (Mott-VRH) mechanism instead of the simple thermally activated hopping exhibited by the AF memristors. The conductance governed by the Mott-VRH law is given by
\begin{equation}\label{eq:Mott_eq}
    G_{\textrm{Mott}}(T) = G_{0,\textrm{Mott}}T^{-1/2}\exp \left[-\left(\frac{T_{0, \textrm{Mott}}}{T}\right)^{1/4}\right]
\end{equation}
where $T_{0, \textrm{Mott}}$ is a characteristic temperature coefficient and $G_{0, \textrm{Mott}}$ is the conductance at $T_{0}$ \cite{mott2012electronic,Yildiz_2008}. Meanwhile at lower temperatures, $\ln \left(GT\right)$ fits linearly with $T^{-1/2}$ which suggests that the conduction mechanism is described by the ES-VRH law below \SI{77}{\kelvin} (see \cref{fig:conduct}(a)). The ES-VRH governed conductance is given by
\begin{equation}\label{eq:ES_eq}
    G_{\textrm{ES}}(T)=G_{0,\textrm{ES}}T^{-1}\exp \left[-\left(\frac{T_{0, \textrm{ES}}}{T}\right)^{1/2}\right]
\end{equation}
Thus, the total conductance of the CR memristor can be defined using the sum of the contributions from the Mott-VRH conduction mechanism and the ES-VRH conduction mechanism \cite{Yildiz_2008}:
\begin{equation}\label{eq:global_conduct}
    G(T) = G_{\textrm{Mott}}(T) + G_{\textrm{ES}}(T)
\end{equation}
The total conductance predicted by \cref{eq:global_conduct} is in good agreement with our experimental data using the fitted parameters for the ES-VRH and Mott-VRH conduction mechanisms (see \cref{fig:conduct}(c) for the fit and \cref{tab:fit_params} for the fitted parameters).\linebreak

The conduction mechanisms for $T>$\SI{77}{\kelvin} differ for the CR (Mott-VRH) and AF (NNH) memristors which could explain the MIT suppression. The CR memristor exhibits conductances that are two orders of magnitude larger for temperature below the MIT temperature ($\approx$100-\SI{125}{\kelvin}) \cite{Pickett2011}. The conductance of the AF memristor decreases by over two orders of magnitude from \SI{300}{\kelvin} to \SI{4.2}{\kelvin} while it decreases by only a factor of 5 for the CR memristor over the same temperature range. This validates the suppression of the MIT regime for the CR devices, as it does not exhibit an important conductance drop with decreasing temperature. The CR device shows a crossover from $T^{-1/4}$ to $T^{-1/2}$ dependence within the temperature range from \SI{80}{\kelvin} to \SI{100}{\kelvin} as demonstrated in Ref.\cite{Yildiz_2008} for TiO$_2$ thin films.
This significant change in the conduction mechanisms governing the memristor conductance with respect to the temperature suggests that Ti$_4$O$_7$ has been suppressed. One possible explanation is that the previously crystalline conductive filament became amorphous due to a quenching effect during the reforming procedure at cryogenic temperatures, similarly to the switching mechanism of phase-change memory \cite{LeGallo2020}, leading to the creation of an amorphous phase inside the conductive filament. This phase change of the oxide layer is validated by the retention of the reformed state demonstrated in \cref{fig:DC}(d), which suggests a durable suppression of Ti$_4$O$_7$ in the vicinity of the conductive filament. \linebreak

\begin{table}
\caption{\label{tab:fit_params} Fitting parameters for the CR memristor conductance at the LRS (see the blue diamonds in \cref{fig:conduct}(c)) and for the AF memristor conductance at the LRS (see the red squares in \cref{fig:conduct}(c)). N.A. indicates that the fitting parameter is not applicable.}
\begin{ruledtabular}
\begin{tabular}{cccc}
Fit&$G_0$ (\SI{}{\siemens})&$T_0$ (\SI{}{\kelvin})&$E_a$ (\SI{}{\milli\electronvolt})\\
\hline
ES-VRH (no MIT) & $1.15\times 10^{-2}$ & 54.49 & N.A.\\
Mott-VRH (no MIT) & 42.15 & $1.42\times 10^{6}$& N.A.\\
NNH Arrhenius (MIT) & $7.5\times 10^{-4}$& N.A. & 41.8 \\
ES-VRH (MIT) & $3.02\times 10^{-6}$& 46.1 & N.A. \\
\end{tabular}
\end{ruledtabular}
\end{table}

\begin{figure}
\begin{minipage}[b]{1\columnwidth}
  \includegraphics[width=1\linewidth]{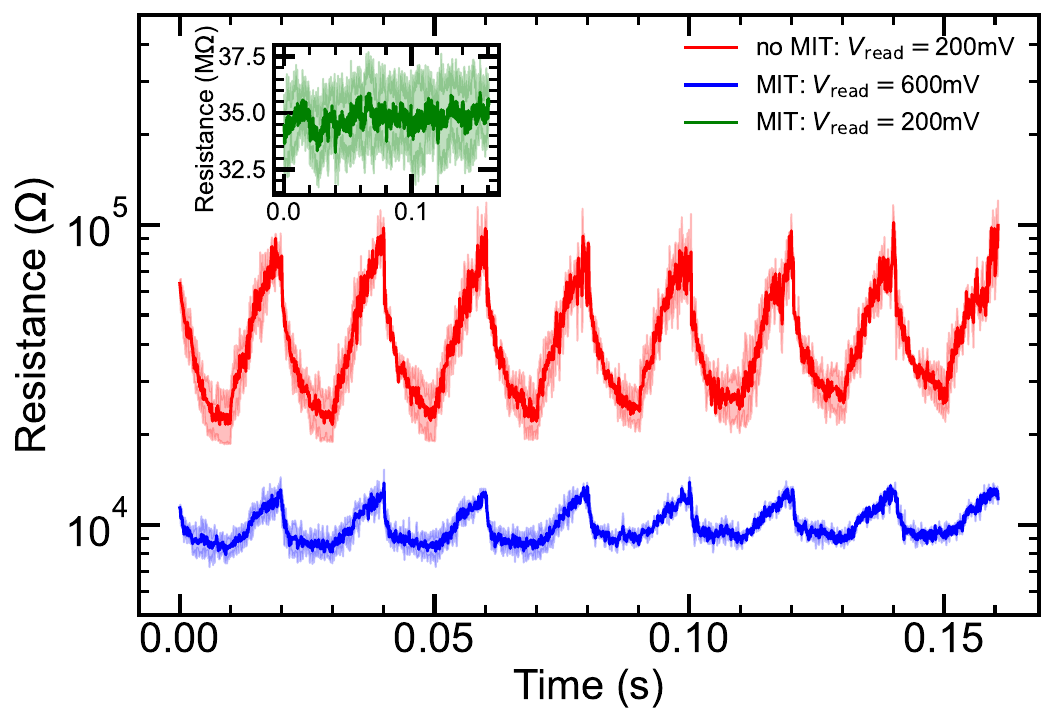}
  \caption{\label{fig:pulsed} Long-term potentiation (LTP) and depression (LTD) curves for a CR memristor (red) and an AF memristor (blue and green) at \SI{4.2}{\kelvin}. The write pulse parameters are $t_{\textrm{write}}=$\SI{200}{\nano\second} and $V_{\textrm{write}}=$ $\pm$\SI{0.9}{\volt}. The read pulse parameters are $t_{\textrm{read}}=$\SI{10}{\micro\second} and $V_{\textrm{read}}=$\SI{0.2}{\volt} or \SI{0.6}{\volt}. Each curve is an average over 10 measurements, and the standard deviation is shown in the background shadow.}
\end{minipage}\hfill 
\end{figure}

Finally, we investigate the impact of the MIT suppression induced by cryogenic reforming on the analog switching dynamics of our Al$_2$O$_3$/TiO$_{\textrm{2-x}}$ memristors using pulse programming. To do this, we measure the same device before and after reforming. Resistance tuning is assessed on this memristor by measuring long-term resistance potentiation (LTP) and depression (LTD) at \SI{4.2}{\kelvin}. A write/verify pulse train comprised of one \SI{200}{\nano\second}/$\pm$\SI{0.9}{\volt} write pulse and one \SI{10}{\micro\second}/\SI{0.2}{\volt} (or \SI{0.6}{\volt}) read pulse is applied 50 times with a constant positive amplitude for the write pulse (LTP) and 50 times with a constant negative amplitude (LTD). This pattern is repeated seven times to verify the stability of the LTD/LTP tuning. This whole protocol is performed 10 times in order to evaluate the cycle-to-cycle variability (background shadow in \cref{fig:pulsed}).\linebreak
For the typical read voltage (\SI{0.2}{\volt}), the resistance of the AF memristor remains unchanged, as the reading is performed in the insulator region of the memristor as depicted in \cref{fig:DC}(d). By increasing the reading voltage to \SI{0.6}{\volt} to operate above the insulator regime, one can observe typical LTD/LTP resistance tuning for the AF memristor which exhibits an on/off ratio of $\approx2$ for a write amplitude of $\pm$\SI{0.9}{\volt}. After reforming we repeat the measurement with the same write amplitude but lowered the read voltage to \SI{0.2}{\volt}. In this case, the CR memristor demonstrates a larger on/off ratio of $\approx5$ in a larger absolute memory window. \linebreak

The LTP and LTD characteristics are quasi-linear and stable for both the AF memristor read at \SI{0.6}{\volt} and CR memristor read at \SI{0.2}{\volt,} which ensures fully analog tunability of the memristor resistance over the respective \SI{7}{\kilo\ohm} to \SI{14}{\kilo\ohm} and \SI{20}{\kilo\ohm} to \SI{100}{\kilo\ohm} ranges before and after the reforming. Thus, reforming the memristors increases the resistance on/off ratio, which allows a larger resistance tunability in cryogenic neuromorphic circuit and an increased resolution for synaptic weight encoding. Additionally, one can observe that the reading region for CR memristors at \SI{4.2}{\kelvin} (0-\SI{0.6}{\volt}) is similar to the one exhibited at room temperature (see \cref{fig:DC}(c)). Having a large read region is an important feature for vector matrix multiplication (VMM) \cite{Li2017}, as it improves the achievable resolution for the input vector and thus provides a more accurate VMM result overall \cite{Yao2020}. \linebreak

On the other hand, it was observed that AF memristors offered a smaller read region, ranging from \SI{0.6}{\volt} to \SI{0.8}{\volt}, along with smaller resistance values when the AF memristor is read in this region. Both the smaller resistance and the higher read voltage range introduce additional power consumption during inference which is detrimental in the context of cryogenic environments. However in the insulator region, the conductance of the AF memristor is not changed i.e., for amplitudes lower than \SI{0.6}{\volt}. This property can be leveraged and used as a selector \cite{Beilliard2020}. It would allow one to  accurately program each device in a crossbar without influencing the resistance of the surrounding memristors \cite{Yan2016}. This presents an interesting trade-off between the programming accuracy of the weights, the power consumption and the on/off resistance ratios.
For applications that do not require to reprogram the crossbar several times, such as neural network inference accelerators \cite{Czischek2021},  the CR approach should be preferred by taking advantage of the larger on/off ratios. For programming-intensive applications, the AF approach should be considered as the MIT induced selector could allow to program accurately large crossbars.\linebreak

In conclusion, we conducted electrical characterizations of Al$_2$O$_3$/TiO$_{2-x}$ memristors at cryogenic temperatures and proposed a method called cryogenic reforming to suppress the metal-insulator transition introduced by Ti$_4$O$_7$ below \SI{125}{\kelvin}. Using the cryogenic reforming methodology, we demonstrated resistive switching at \SI{4.2}{\kelvin} without insulator regime and a retention of the MIT suppression for over a month hinting at a permanent change of the conductive filament. Moreover, the study of temperature dependence of the conductance suggests different conduction mechanisms for the AF and CR memristors from \SI{77}{\kelvin} to \SI{200}{\kelvin}. This validates the idea that the conductive filament has undergone intrinsic changes through the reforming process. The CR memristor exhibits a conduction drop at \SI{4.2}{\kelvin} that is 100 times smaller than that of the AF memristor due to the MIT suppression. Finally, we demonstrated that CR memristors have a fully analog behavior at \SI{4.2}{\kelvin} which would enable the development of memristor-based  cryogenic electronics applied to quantum technologies. AF devices exhibit a smaller memory window but have a built-in selector below \SI{100}{\kelvin} thanks to the MIT that could be utilized for large crossbar applications.
Our work is a stepping stone towards memristor-based cryogenic applications in quantum technologies showing the viability of the TiO$_{\textrm{2-x}}$-based memristor at deep cryogenic temperatures. \linebreak

This work was supported by the Natural Sciences and Engineering Research Council of
Canada (NSERC) and the Canada First Research Excellence Fund (APOGEE). This research was undertaken thanks to support from Laboratoire Nanotechnologies Nanosystèmes (LN2), which is a French–Canadian Joint International Research Laboratory (IRL3463), funded and co-operated by Centre National de la Recherche Scientifique (CNRS), Université de Sherbrooke, Université de Grenoble Alpes (UGA), École Centrale Lyon (ECL), and Institut National des Sciences Appliquées (INSA) Lyon. We acknowledge financial support from the Fonds de Recherche du Québec Nature et Technologie (FRQNT).

\section*{AUTHOR DECLARATIONS}
\RaggedRight\noindent\textbf{Conflicts of Interest}\linebreak
The authors have no conflicts to disclose.\linebreak
\textbf{Author Contributions}\linebreak
\justifying \textbf{Pierre-Antoine Mouny}: Conceptualization (equal); Data curation (equal); Formal analysis (lead); Investigation (lead); Methodology (lead); Validation (equal); Visualization (lead); Writing – original draft (lead); Writing – review \& editing (equal). \textbf{Raphaël Dawant}: Conceptualization (equal); Data curation (supporting); Investigation (supporting); Resources (lead); Validation (equal); Visualization (supporting); Writing – original draft (supporting); Writing – review \& editing (equal). \textbf{Bastien Galaup}: Data curation (equal); Formal analysis (supporting); Investigation (supporting); Methodology (supporting); Validation (equal); Writing – review \& editing (equal). \textbf{Serge Ecoffey}: Project administration (supporting); Resources (supporting); Supervision (supporting); Validation (equal); Writing – review \& editing (equal) \textbf{Michel Pioro-Ladrière}: Project administration (supporting); Funding Acquisition (supporting); Supervision (supporting); Writing – review \& editing (equal).  \textbf{Yann Beilliard}:  Conceptualization (equal); Project administration (lead); Investigation (supporting); Supervision (supporting);  Validation (equal);  Writing – original draft (supporting); Writing – review \& editing (equal) and \textbf{Dominique Drouin}: Conceptualization (equal); Funding acquisition (lead); Project administration (supporting); Supervision (lead);  Validation (equal);  Writing – original draft (supporting); Writing – review \& editing (equal)\linebreak
\RaggedRight\textbf{DATA AVAILABILITY}\linebreak
\justifying\indent The data underlying the results presented in this paper are not publicly available at this time but may be obtained from the authors upon reasonable request.\linebreak

\noindent \textbf{REFERENCES}

\nocite{*}
\bibliography{aipsamp}

\end{document}